\documentclass[amsmath, amssymb, aps, pra,reprint, twocolumn,longbibliography,nofootinbib]{revtex4}

\usepackage[utf8]{inputenc}
\usepackage{braket}
\usepackage{graphicx}

\begin{document}

\title{Parallel Hybrid Networks:\\ an interplay between quantum and classical neural networks}

\author{Mohammad Kordzanganeh}
\author{Daria Kosichkina}
\author{Alexey Melnikov}
\thanks{Corresponding author, e-mail: alexey@melnikov.info
\begin{center}
\fbox{
\begin{minipage}{0.45\textwidth}
Please check the published version, which includes all the latest additions and corrections: Intell. Comput. 2:0028, 2023, DOI: \href{https://doi.org/10.34133/icomputing.0028}{10.34133/icomputing.0028}
\end{minipage}
}
\end{center}
}
\affiliation{Terra Quantum AG, Kornhausstrasse 25, 9000 St.~Gallen, Switzerland}

\begin{abstract}
Quantum neural networks represent a new machine learning paradigm that has recently attracted much attention due to its potential promise. Under certain conditions, these models approximate the distribution of their dataset with a truncated Fourier series. The trigonometric nature of this fit could result in angle-embedded quantum neural networks struggling to fit the non-harmonic features in a given dataset. Moreover, the interpretability of neural networks remains a challenge. In this work, we introduce a new, interpretable class of hybrid quantum neural networks that pass the inputs of the dataset in parallel to 1) a classical multi-layered perceptron and 2) a variational quantum circuit, and then the outputs of the two are linearly combined. We observe that the quantum neural network creates a smooth sinusoidal foundation base on the training set, and then the classical perceptrons fill the non-harmonic gaps in the landscape. We demonstrate this claim on two synthetic datasets sampled from periodic distributions with added protrusions as noise. The training results indicate that the parallel hybrid network architecture could improve the solution optimality on periodic datasets with additional noise.  
\end{abstract}

\maketitle

\section{Introduction}
Machine learning and quantum computing have become attractive research areas in recent years. The quest for an efficient quantum neural network (QNN) has dominated the cross-section of these two technologies. Many suggestions have been made for the potential inner workings of a classically-intractable quantum machine learning model \cite{rig-rob,caro,challengesandopportunities,review_paper}, but theoretical and hardware limitations could prove challenging to implement. The noise-free barren plateau problem \cite{bp} or the curse of dimensionality \cite{power-data} are examples of the theoretical challenges with QNNs. At the same time, the hardware limitations point to the industry limits on the accuracy, and the number of qubits \cite{benchmarking,noise-bp}.  
Therefore, contemporary practical use of quantum technologies in machine learning should come from complementary quantum-classical architectures, called hybrid quantum neural networks (HQNN), that employ relatively small, realisable quantum circuits and classical multi-layered perceptrons (MLP) where the two work in tandem.  The works in \cite{asel-paper1,asel-paper2,asel-paper3,thales,bert,class2quant} explored the applicability and performance of sequential HQNNs, where MLPs and QNNs are connected in series, passing the information from one network to another. The sequential HQNNs could introduce information bottlenecks in the representational power of the model, which could limit the expressivity of the network. This work explores the theoretical basis of parallel HQNNs, where variational quantum circuits (VQC) and MLPs process information in parallel. The approach is based on the universality theorems from two sources: 1) MLPs can produce non-harmonic functions \cite{nn-book} and  2) QNNs fit smooth truncated Fourier series on the training data \cite{schuld-fourier}.  This work was inspired by the Fourier neural operator introduced by Ref.~\cite{fno_paper}.

In Sec \ref{sec:literature-review}, we review the theoretical foundations of MLP and VQCs, and in Sec \ref{sec:PHN}, we introduce the design and experimental results of PHN. In Sec \ref{sec:primacy}, we address the potential problem of component primacy in training PHNs, where either the VQC or the MLP could dominate the training, and propose a remedy to it. Finally, in Sec \ref{sec:conclusion}, we summarise our findings and discuss future directions.

\section{Theoretical foundation} \label{sec:literature-review}

In this study, we concentrate on solving a supervised regression problem using a dataset $(\mathbf{x}_i,y_i)$, where $\mathbf{x}_i \in \mathbf{\mathcal{X}}$ is a feature vector and $y_i \in \mathcal{Y}$ is the label. We aim to discover a function $f(\mathbf{x})$ that can approximate the labels $y$ of out-of-sample features. To achieve this, we create a machine learning model with parameters $\mathbf{\theta}$ to create the functionality, $f_{\mathbf{\theta}}(\mathbf{x})$. We adjust these parameters according to the training sample to maximise the probability of obtaining the correct label for a given feature. The general functionality $f$ is a machine learning architecture, while a specific realisation of its parameters, $\mathbf{\theta}$, is a machine learning model. In the subsequent sections, we will explore two well-known architectures and then use that theoretical foundation to justify the PHN architecture in Sec \ref{sec:PHN}.
\subsection{Multi-layered perceptrons}

MLPs constitute a large class of successful machine learning architectures. They are directional graphs whose nodes are ordered in one-dimensional layers which take input from the previous layer and provide the next layer with their outputs. In the case of fully connected MLPs (FCN), all neurons of each layer feed information to all the neurons in their immediate front neighbourhood. Each edge of the graph has an associated multiplicative factor (weight), and each neuron has an associated additive quantity (bias), which together form the parameters of the MLP. The first neural layer is called the input layer, and the last is the logits. Ref.~\cite{bishop} provides a comprehensive overview of neural networks and their properties. 

Ref.~\cite{cybenko} showed that MLPs are asymptotically universal approximators whose fit on the training data becomes perfect as the numbers of neurons in the intermediate neural layers approach infinity.  Moreover, \cite{hornik,nn-book} proved this using a novel graphical method that showed a fully-connected network with a single intermediate neural layer can approximate any function by fitting a superposition of rectangular waves. To utilise this graph as a machine learning architecture, we could encode the features of a data point taken from the sample, $\mathbf{x}_i$, onto the input layer and then propagate their values through the graph by multiplying their values by the weights of the architecture $\mathbf{w}$ and adding the biases $\mathbf{b}$

\begin{equation} \label{eqn:mlp}
    h_i = \sigma(w^{(0)}_{i,j} x_j + b_i),
\end{equation}
where $\sigma$ is known as the activation function\footnote{In this work, the particular functionality of the activation function is not material, and for simplicity, we take it to be the sigmoid function everywhere, $\sigma(t) = \frac{1}{1+e^{-t}}$, and otherwise only where clearly stated.}, $h_i$ indicates the values of the consecutive neural layer, and Einstein's summation notation is implied.  The propagation process can be passed along to the entire graph until we arrive at the terminal nodes, the prediction of the MLP. 

\subsection{Variational quantum circuits} \label{sec:vqc}

\begin{figure*}[!htpb]
    \centering
    \includegraphics[width=0.7\linewidth]{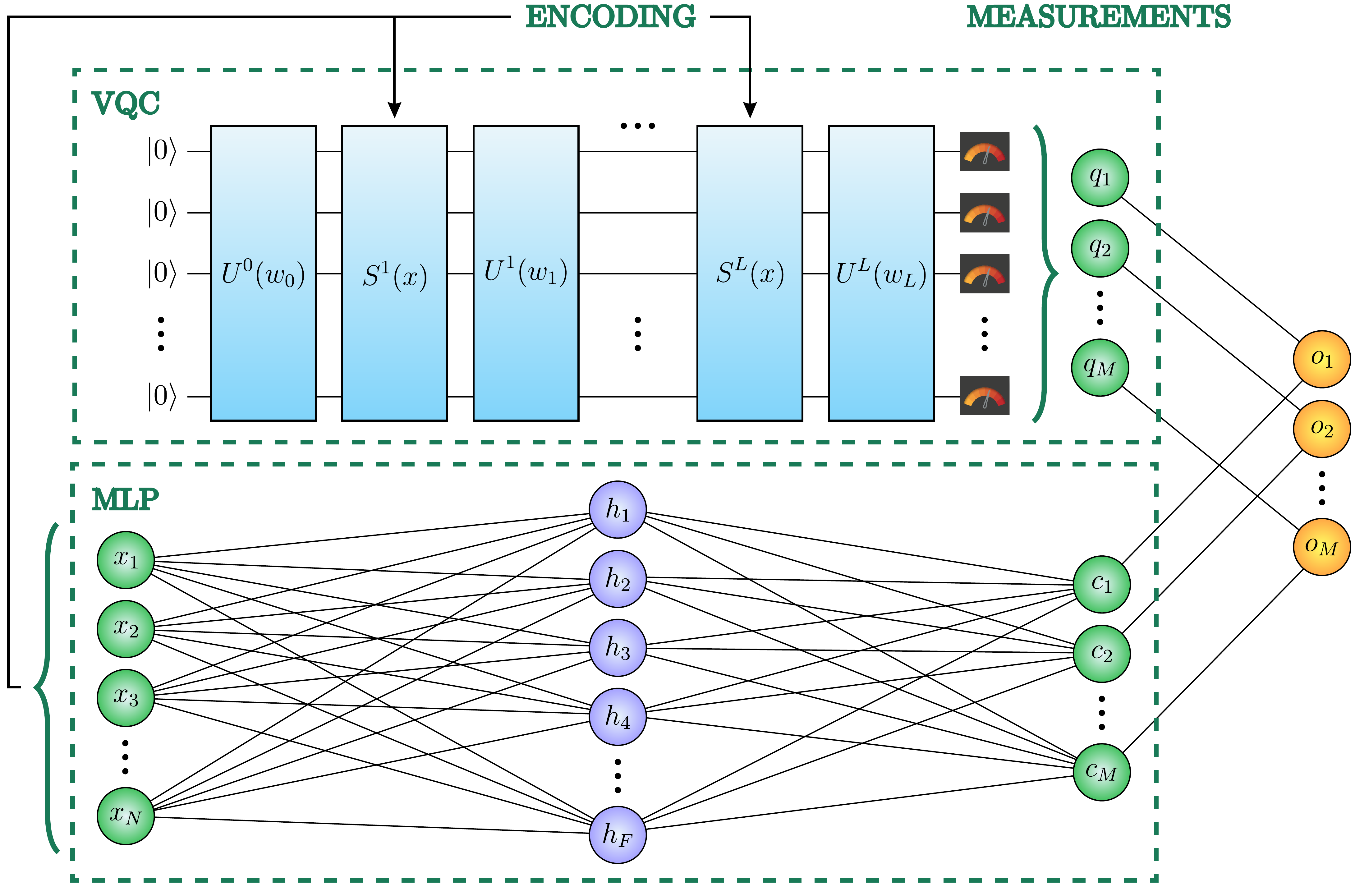}
    \caption{The general architecture of the PHN. The PHN takes an input vector of features and passes them to an angle-embedded VQC (a VQC that uses single-qubit, Pauli gate embedding of features without applying any non-linear kernel on the features) of the appropriate architecture in parallel to a multi-layered perceptron with a single hidden layer of the appropriate size.  The outputs of the VQC are then combined linearly with the outputs of the MLP to produce a final output vector.}
    \label{fig:phn_general}
\end{figure*}

Variational quantum circuits (VQC) employ variational group rotations to create a machine learning architecture on a quantum computer \cite{Peruzzo,review_paper,benedetti-paper,schuld-book,errorperf}. To construct a VQC, one could start by creating a quantum node of several qubits in the ground state.  Then, a series of variational and fixed quantum gates can be applied to the circuit.  The variational gates could include the Pauli rotation gates, which require single-qubit time evolution Hamiltonians of the respective Pauli gate.  The fixed quantum gates might consist of the controlled-NOT (CNOT) and the Hadamard (H) gates. We split the variational gates into embedding, and trainable gates, which encode the features, $\mathbf{x}$, and act as model parameters, $\mathbf{\theta}$.  At the end of the circuit, we measure the qubits in a specified basis, such as the Pauli bases, and obtain either a $0$ or a $1$. After many iterations of the circuit, we can find the likelihood of getting a $0$ over $1$, and by taking the average, we can obtain the expectation value of the circuit.  By the Born rule \cite{born}, we can find this probability by taking the expectation of the measurement matrix, $M$, and then using it as the output of our model:

\begin{equation}
    f(\mathbf{x},\mathbf{\theta}) = \bra{\psi(\mathbf{x},\mathbf{\theta})}M\ket{\psi(\mathbf{x},\mathbf{\theta})},
\end{equation}
where $\ket{\psi(\mathbf{x},\mathbf{\theta})}$ denotes the state of the quantum circuit before the measurement.  We can improve this approximation to the labels by optimising the parameters of the VQC, $\mathbf{\theta}$. \cite{schuld-fourier} proved that VQCs are also universal approximators, and the way they work is by fitting a truncated Fourier series over the samples:
\begin{equation}
    f(x) = \sum_{k=-L}^{L} c_k e^{ikx},
\end{equation}
where L is the highest degree Fourier term expressible by the VQC.  

In the functioning of VQCs, the interaction between the variational and fixed quantum gates forms a critical aspect. The variational gates navigate the quantum system through a sequence of transformations, manipulating quantum states based on the data $x$ and trainable parameters $\mathbf{\theta}$. On the other hand, fixed quantum gates, such as CNOT and Hadamard gates, exhibit predictable behaviours and serve to entangle and manipulate qubits. While the fixed gates maintain system structure, the variational gates enhance adaptability, thereby enabling the circuit to learn and adapt to new data.

\section{Results -- parallel hybrid networks} \label{sec:PHN}

We split the HQNN hybrid interfaces into two categories: 1) sequential: where the classical and quantum parts feed directly into each other, and 2) parallel: where a classical multi-layered perceptron and a variational quantum circuit in parallel process the same information. In this section, we take an in-depth look into HQNNs of the latter type and the functions they represent. Appendix~\ref{sec:info_bottleneck} provides an empirical comparison between the two categories, but in the following, we explore the latter type. We shall refer to these networks as parallel hybrid networks (PHN). Fig.~\ref{fig:phn_general} shows the general architecture of PHNs. The combination is a weighted linear addition with trainable weights. These weights determine the contribution of each network to the final output. The specific VQC used here is a generalised data re-uploading VQC, where $K$ qubits are initialised in the state $\ket{0}^{\otimes K}$. Then in alternation, a series of variational and encoding layers are applied.  The encoding layers $S$ take the input features, $\{x_1,\cdots,x_N\}$, and encode them in a unitary transformation which is then applied to the state of the qubit.  The variational layers, $U$, are unitaries that encapsulate the VQC model parameters as an operator that can be used for the quantum state of the network. Finally, the measurements are where the quantum information collapses into $M$ classical outputs, which can be obtained by taking the expectation value of the circuit with respect to the measurement observable.  Note the difference between $M$, the number of classical outputs out of the VQC, and $K$, the number of qubits, and that they are not necessarily the same, as often we are only required to measure some of the qubits. 

\begin{figure*}[!tbph]
  \centering
    \includegraphics[width=\textwidth]{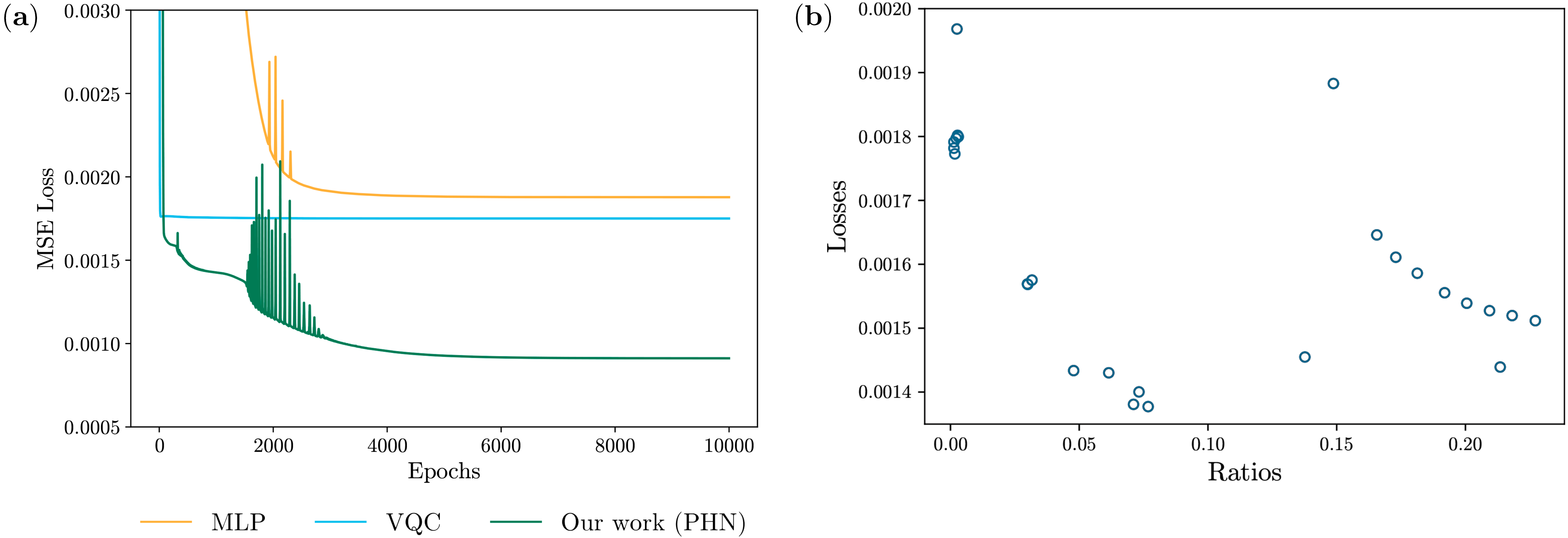}
    \caption{(a) The training losses of the individual elements of the PHN when trained separately, as well as the full PHN.
     (b) The scatter plot of final losses and their respective final ratios after 1000 training epochs. The optimal loss value is achieved at non-zero ratios, where ratios to the side of this value provide sub-optimal losses.  Note that this figure only includes the runs with learning rates whose final loss is low enough for comparison.}
     \label{fig:loss}
\end{figure*}

\begin{figure}[ht]
    \centering
    \includegraphics[width=1.0\linewidth]{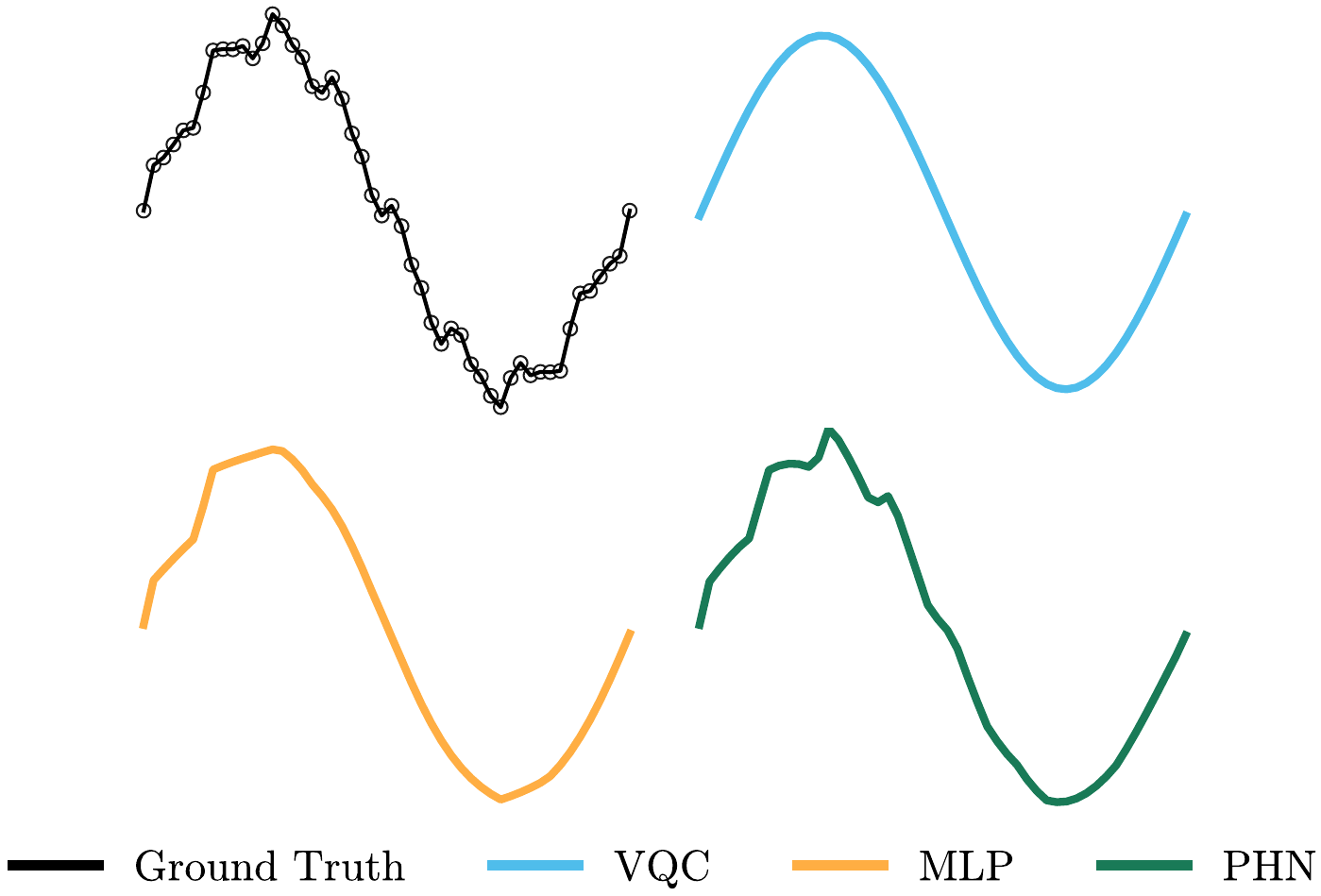}
    \caption{The functional fits of each architecture to the ground truth. The VQC, expectedly, produced a sinusoidal curve. The MLP created an overall curve close to a sinusoidal curve but with jagged edges.  The PHN, however, produced the best result, predicting the protrusion at the peak.}
    \label{fig:fits} 
\end{figure}

In parallel, the fully connected MLP also takes in the $N$ features and passes them to a single layer of hidden neurons of size $F$ by multiplying the feature vector by a weight matrix of size $N\times F$. Then, biases are applied to these values and scaled using an activation function.  An activation function is necessary for adding non-linearity to an otherwise linear system.  Then, the neurons are propagated to $M$ MLP output neurons with their own biases and activation functions, denoted as $\{c_1,\cdots,c_M\}$.  The MLP and VQC outputs are then combined, using a two-to-one linear weight layer, to form the PHN outputs, $\{o_1,\cdots,o_M\}$.  This final layer combines the first output of the VQC with the first output of the MLP: $o_1 = s^c_1 c_1 + s^q_1 q_1$, and similarly for all the $M$ outputs, where $(\{s^q\},\{s^c\})$ are trainable parameters.

In Sec \ref{sec:literature-review}, we saw that an MLP with a single hidden layer created a non-harmonic functional fit for the dataset and that the VQC created a truncated Fourier series, a harmonic function.  Thus, a network combining these results could map the smooth, sinusoidal parts through the VQC and fill the protruding sections via the MLP.   This complementary setting has the potential to approximate a function that fits the dataset both in the position space (MLP) and in the conjugate momentum space (VQC). We could compare this duality to the Fourier neural operator in Ref.~\cite{fno_paper} or the models with benign overfitting in Ref.~\cite{evanpeters}. The scope of this work includes architectures that use multiple VQCs (MLPs) in parallel, as they can always be combined to form a single VQC (MLP). 

\subsection{Performance}\label{sec:experiments}
We start with a ground truth consisting of an overall single-frequency sinusoidal function and then introduce high-frequency perturbations to this system. Specifically, the functional form was  

\begin{align*}
f = \sin(x) + 0.05 \sin(8x) + 0.03 \sin(16x) + 0.01 \sin(32x),
\end{align*}
which was scaled to -1 and 1. 100 equally-spaced data samples were taken from this distribution for training.  We train a simple PHN, described in detail in Appendix~\ref{sec:setup} with the exact structure of the MLP and the VQC, to recreate the ground truth as accurately as possible. We then train the individual constituents of the same PHN architecture to see their performances.  Fig.~\ref{fig:loss} shows the training loss curves, and Fig.~\ref{fig:fits} shows the best fits that each architecture created for the ground truth. The PHN trains to a lower MSE training loss than its elements, which suggests that adding the VQC improves the overall expressivity of the MLP. Furthermore, by examining the loss curves, we see that the PHN inherits the same speedy descent as the VQC but also shares many of the features present in the MLP loss curve, such as the spikes or the gradual flattening near the end of training. 
  
We see that the PHN outperforms both individual components, which means that both the VQC and MLP contribute to the training, and neither becomes redundant.  In Sec \ref{sec:primacy}, we explore how to measure the contribution of each and how the tuning of hyper-parameters could change this contribution. Moreover, Appendix~\ref{sec:doubled} explores the fairness of this comparison, Appendix~\ref{sec:high-freq} tests the case where the higher frequency term has the larger amplitude, and Appendix~\ref{sec:generalisation} investigates the generalisation ability of the PHN on this dataset.

\begin{figure}[!h]
    \centering
    \includegraphics[width=1.\linewidth]{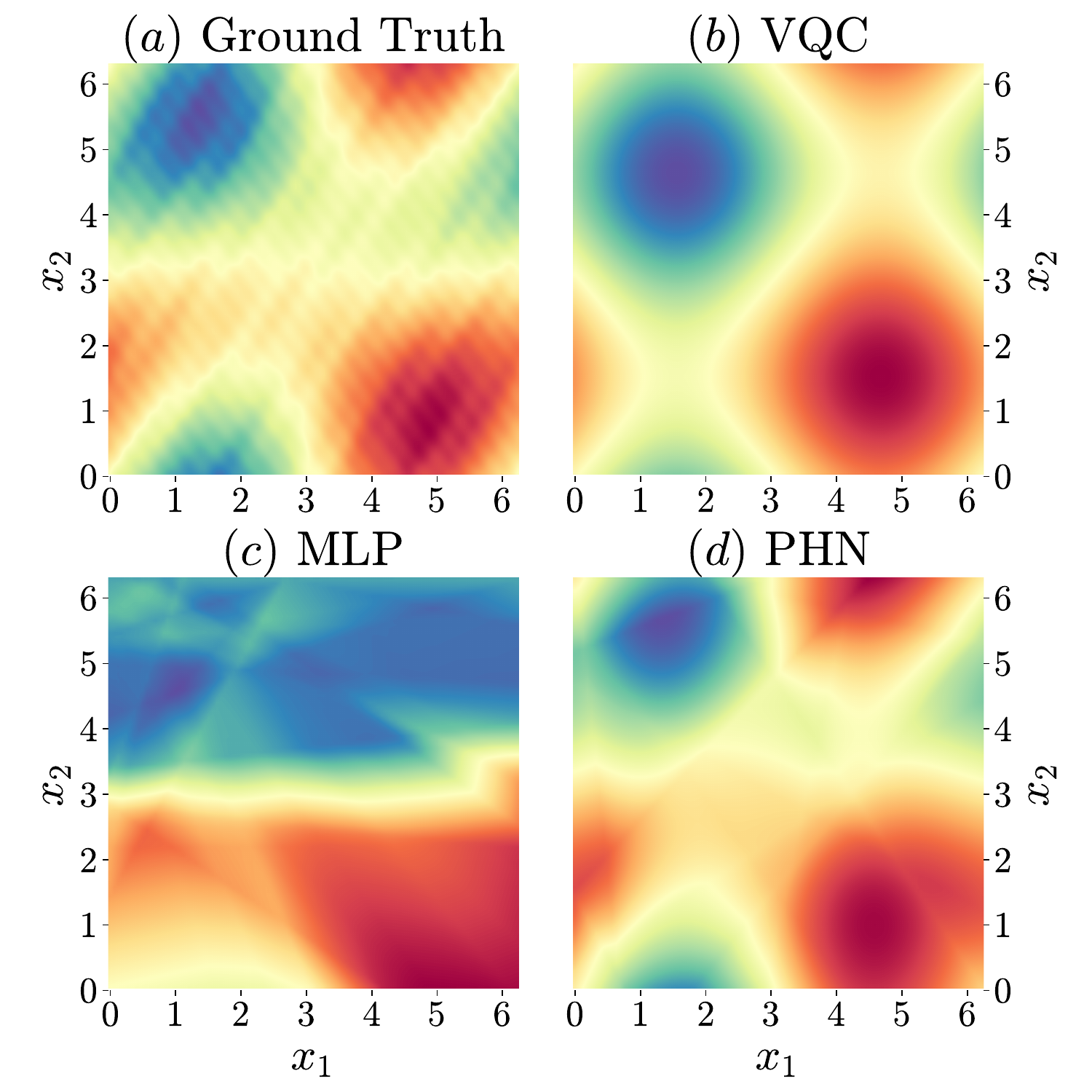}
    \caption{(\textbf{a}) shows the ground truth for our 2D problem in the form of a contour map. The predictions shown are for the VQC (\textbf{b}), MLP (\textbf{c}), and PHN (\textbf{d}). We see that the prediction of the VQC is smooth and convex, whereas the MLP creates jagged shapes. Taking advantage of both of these properties, the PHN represents the harmonic functions of the VQC with the added, necessary protrusions.}
    \label{fig:predictions_2d}
\end{figure}

\begin{figure}[!h]
    \centering
    \includegraphics[width=1\linewidth]{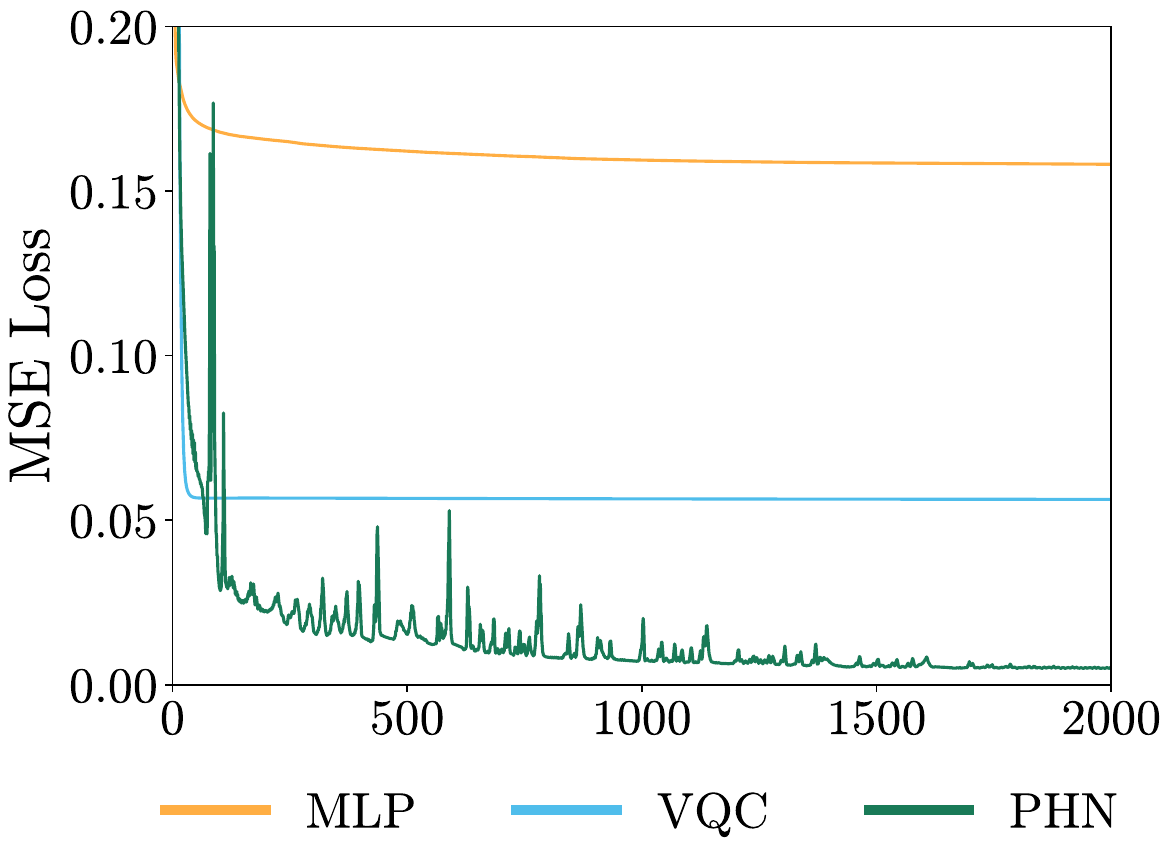}
    \caption{The evolution of the loss function of the PHN model in Fig.~\ref{fig:phn_particular}(b) and its constituents (VQC and MLP) on the 2D dataset.  The PHN achieves an impressively low training MSE loss.}
    \label{fig:loss_2d}
\end{figure}

\subsection{PHN primacy} \label{sec:primacy}
A way to understand the relative contributions of each network is by inspecting the weights of the combination phase, $s_q$ and $s_c$, for VQC and MLP, respectively. In this section, we look at how this contribution can unfold when training the PHN. 

When training a PHN, we must be wary that the VQC and MLP train at different rates.  We define the primacy of one of the constituent architectures (VQC or MLP) over another as when the last weights preceding the PHN output layer vanish for one of the components. Equivalently, this makes the output of the latter network independent from the input features, which would mean that the prediction curve is solely constructed by either the MLP or the VQC. A primacy of this type could prevent the PHN from reaching the global minimum, as it is limited to what only one of the components could offer.

The ratio of the final weights, $r = \frac{\lvert s_c \rvert}{\lvert s_q \rvert}$, was used to track intervals of different primacy regimes recorded for different hyper-parameterisations of the PHN. Specifically, we fixed the learning rate of the VQC at 0.01 and then selected the learning rate of the MLP from 54 values of $\alpha_c \in \{1.0\mathrm{e}{-7},2.0\mathrm{e}{-7},\cdots,9.0\mathrm{e}{-7},1.0\mathrm{e}{-6},\cdots,9.0\mathrm{e}{-2}\}$. We, then, trained the PHN for 1000 epochs at a fixed initialisation point. Lastly, we recorded the ratios $r$ throughout each training. The bigger the ratio, the more the MLP would contribute compared with the VQC. Notably, even a small contribution could make a critical difference, and primacy occurs only when one of the contributions completely vanishes. 

The dependence of the final loss on the ratio of the final weights for the VQC and MLP, shown in Fig.~\ref{fig:loss}(b), highlights the potential for either component to dominate training in the PHN architecture. The results exhibit an optimal range for the ratio between 0.1 and 1, indicating that a balanced contribution from the VQC and MLP is desirable for achieving the best results. It is also evident that complete MLP primacy, where the ratio approaches $0$, leads to worse final losses. However, we also observe that adjusting the learning rates of the two components can sometimes improve the loss. Therefore, tuning the learning rates of the VQC and MLP is crucial to achieve a balanced contribution from both parts and to prevent either component from dominating the training.

\subsection{Scalability and generalisation}
To show the scalability of the PHN, in this section, we try a 2-dimensional problem with the view that this can be scaled to an arbitrarily complex problem with many qubits.  To solve the problem in Fig.~\ref{fig:predictions_2d}(a), a simple PHN, described in Appendix.~\ref{sec:setup}, was employed. The distribution used to create this ground truth was
\begin{align*}
    f(x_1,x_2) = \sin(x_1)+\sin(x_2)+ 0.8\sin(x_1+x_2)\\+0.3\sin(x_1-x_2)  +0.09\sin(8x_1+4x_2)\\ + 0.05\sin(16x_1-12x_2)+0.04\sin(12x_1+8x_2).
\end{align*}
 Note that this function (similarly to the 1D case) was chosen entirely at random to have a coarse harmonic structure (first four terms) as well as high-frequency noise (the last three terms) and not engineered to showcase the PHN in a favourable light. $100$ equidistant points were sampled from this ground truth to create a training set. This set was then trained on only the VQC, MLP, and the complete PHN for $10,000$ epochs. The trained models were then tested on $10,000$ equidistant points data points to see the generalisation ability of each architecture. Figs.~\ref{fig:predictions_2d}(b), (c), and (d) respectively showcase the fit of the VQC, MLP, and PHN, and Fig.~\ref{fig:loss_2d} shows the evolution of their training loss.  The VQC creates a symmetric, sinusoidal pattern, whereas the MLP creates jagged regions to fit the ground truth. However, the PHN can generalise the ground truth by employing both elements and thus creates a closer fit, which could mean that for such datasets, the PHN could provide a high generalisation power over the MLP or the VQC. 

\section{Conclusion} \label{sec:conclusion}

Overall, our findings demonstrate the potential of PHN as a powerful tool for quantum machine learning. It is a hybrid architecture that can extract harmonic and non-harmonic features from a dataset. By leveraging its unique architecture, the PHN can learn complex patterns and relationships within the data that might be difficult to capture using traditional machine learning algorithms.

However, it is essential to note that the performance of the PHN is highly dependent on the choice of hyperparameters. The number of layers, neurons in each layer, activation functions, and learning rate are crucial in determining how well the network performs on a given task. Therefore, hyperparameter tuning is a critical step in training a successful PHN. One potential direction for future research is to explore using a custom learning rate scheduler to modify the learning rate during training. A learning rate scheduler can dynamically adjust the learning rate based on the network's performance on the training set, allowing the model to learn more efficiently and converge faster. Implementing a learning rate scheduler may further improve the performance of the PHN on a wide range of tasks.

\bibliography{main}
\bibliographystyle{unsrt}
\newpage

\appendix 
\clearpage
\newpage

\section{The exact experimental setup}\label{sec:setup}

\begin{figure*}[t]
    \centering
    \includegraphics[width=0.8\linewidth]{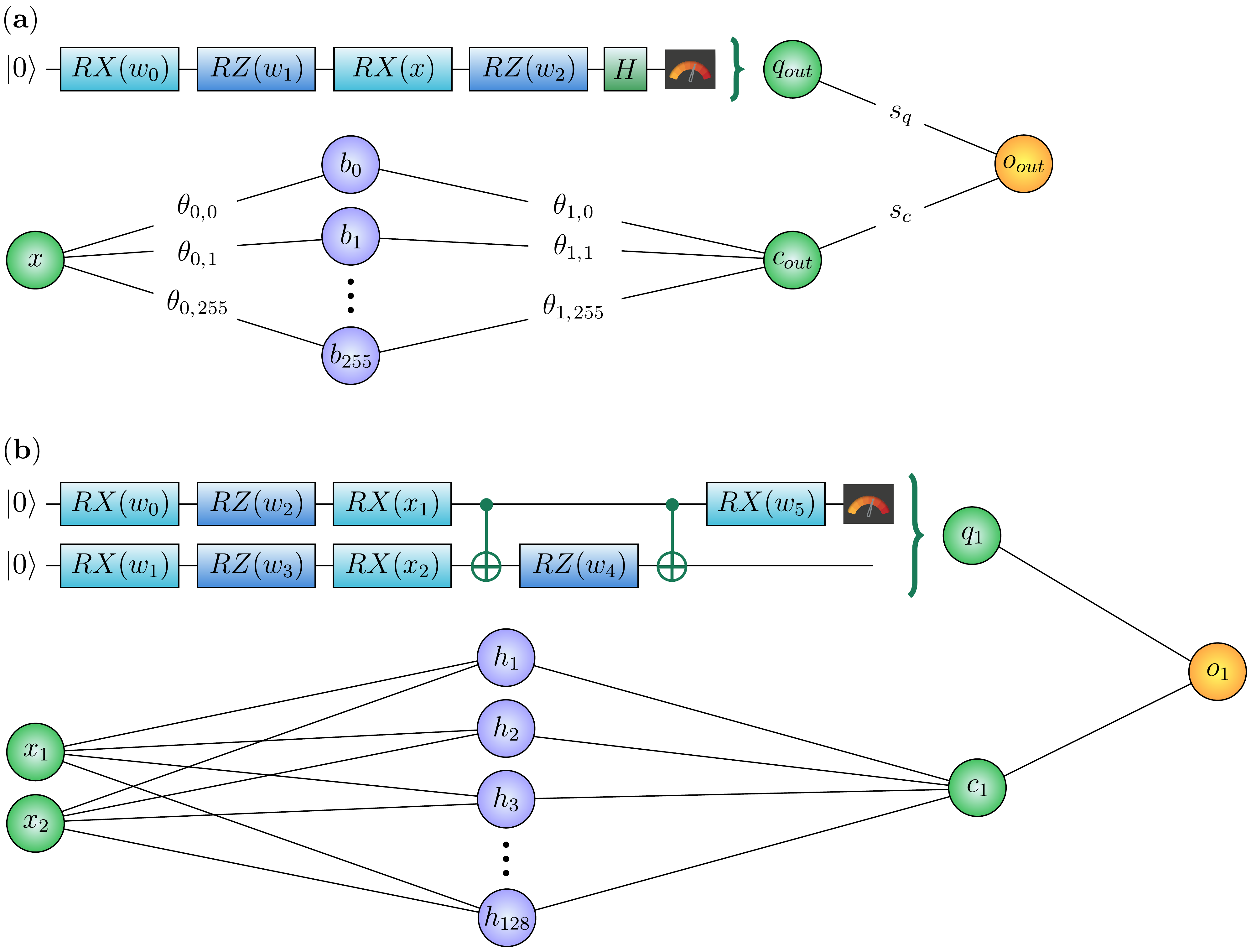}
    \caption{(a) The schematic diagram describes the joint work of the VQC with one qubit and a basic MLP architecture. (b) The architecture includes a two-qubit VQC with only a measurement applied to the first qubit and an MLP with 128 neurons in its singular hidden layer.}
    \label{fig:phn_particular}
\end{figure*}

Figs~\ref{fig:phn_particular}(a) and ~\ref{fig:phn_particular}(b) illustrate the PHN example architectures used to produce the 1D and 2D results, respectively. In both cases, the quantum measurement in $q_\mathrm{out}$ was made in the $Z$-basis, and the MLP utilised a single hidden layer with the rectified linear unit (ReLU) and sigmoid activation functions for the hidden and output layers, $c_\mathrm{out}$, respectively. The MLPs were fully connected and included weights and biases with 256 and 128 neurons, respectively. The outputs of the MLP and VQC were linearly combined after being weighted by $s_c$ and $s_q$, respectively. The MLP had 769, the VQC had 3, and the final weighing layer had two parameters.

Fig~\ref{fig:phn_particular}(b) depicts a simple 2-dimensional PHN used to demonstrate the scalability of the PHN. The activation layers employed in the MLP were ReLU and sigmoid for the first and second layers. The VQC produced a single output, $q_1$, which resulted from measuring the state of the VQC in the $Z\bigotimes I$ basis, where the identity measurement $I$ is excluded from the diagram. A learning rate of 0.01 was used for the VQC parameters and 0.001 for all others. We also utilised the Adam optimiser and a learning rate scheduler that multiplied all learning rates by $\gamma=0.99$ every ten epochs.

\section{Information bottlenecks in sequential hybrid networks}\label{sec:info_bottleneck}

This section focuses on the information bottlenecks in sequential hybrid networks, which are the primary motivation behind the invention of the PHN. According to Ref.~\cite{schuld-fourier}, an angle-embedded VQC, such as the ones in Fig.~\ref{fig:phn_particular}, produces a truncated Fourier series of the feature-set to approximate the labels.  Additionally, Ref.~\cite{nn-book} showed that MLPs fit the function in the position space using rectangular protrusions.  In a sequential hybrid network, the information flow depends on the quantum and classical processes sequence. Therefore, information processing is inherently limited by the processing capacities of either VQC or MLP. This represents a bottleneck in the information flow, given that the information output of one process becomes the input for the other.

When the outputs of a VQC, for example, are passed onto an MLP in sequential models, the MLP is limited by the truncated Fourier series of the VQC. This might result in an incomplete or imprecise approximation of the labels, as the MLP's ability to fit the function in the position space may be constrained by the quality of the VQC's output. Similarly, one could reverse the setting, where the output of the MLP is passed onto the VQC. In that case, the VQC might struggle to process the rectangular protrusions provided by the MLP accurately. These limitations in information flow and processing capabilities are what we refer to as the information bottleneck in sequential hybrid networks.

In contrast, parallel hybrid networks (PHN) sidestep these bottlenecks by allowing simultaneous information processing in the quantum and classical domains. Instead of constraining the system by the sequential passage of information, PHN enables more efficient utilisation of both the quantum and classical capabilities. Consequently, the processing power of the PHN is not restricted by the limitations of a single component but instead is governed by the cumulative capacity of all its parts. 

\begin{figure*}
    \centering
    \includegraphics[width=\textwidth]{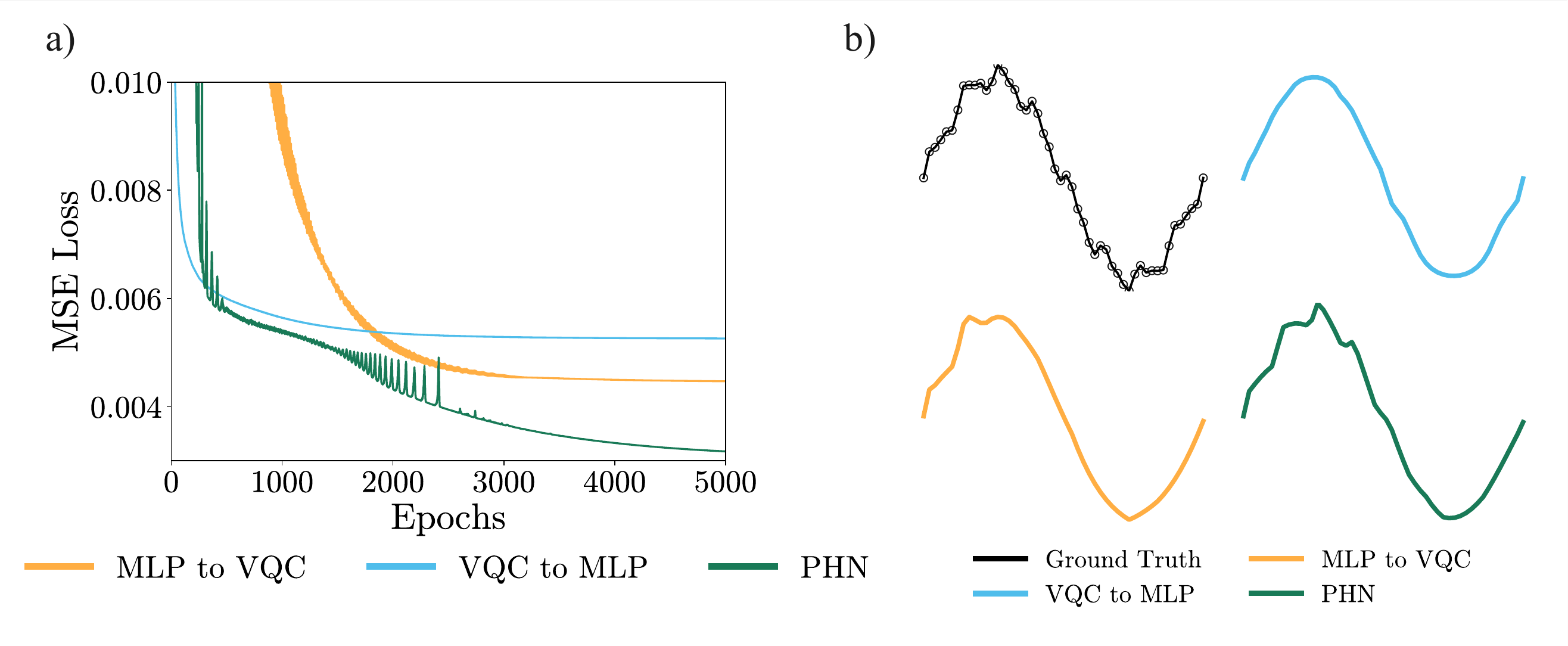}
    \caption{The sequential architectures where in the MLP to VQC structure, the MLP takes in the dataset features and passes its outputs to the VQC.  The output of the VQC is then used as the prediction of the sequential model. The VQC to MLP model performs this operation but in reverse order. In this problem, the PHN clearly has a better fitting ability than the sequential architectures. }
    \label{fig:sequential}
\end{figure*}

Fig.~\ref{fig:sequential} shows the results of sequentially connecting the VQC to the MLP and the MLP to the VQC when trained on the one-dimensional dataset in Sec.~\ref{sec:experiments}.

\section{Comparison with doubled models}\label{sec:doubled}
\begin{figure*}[h]
    \centering
    \includegraphics[width=\textwidth]{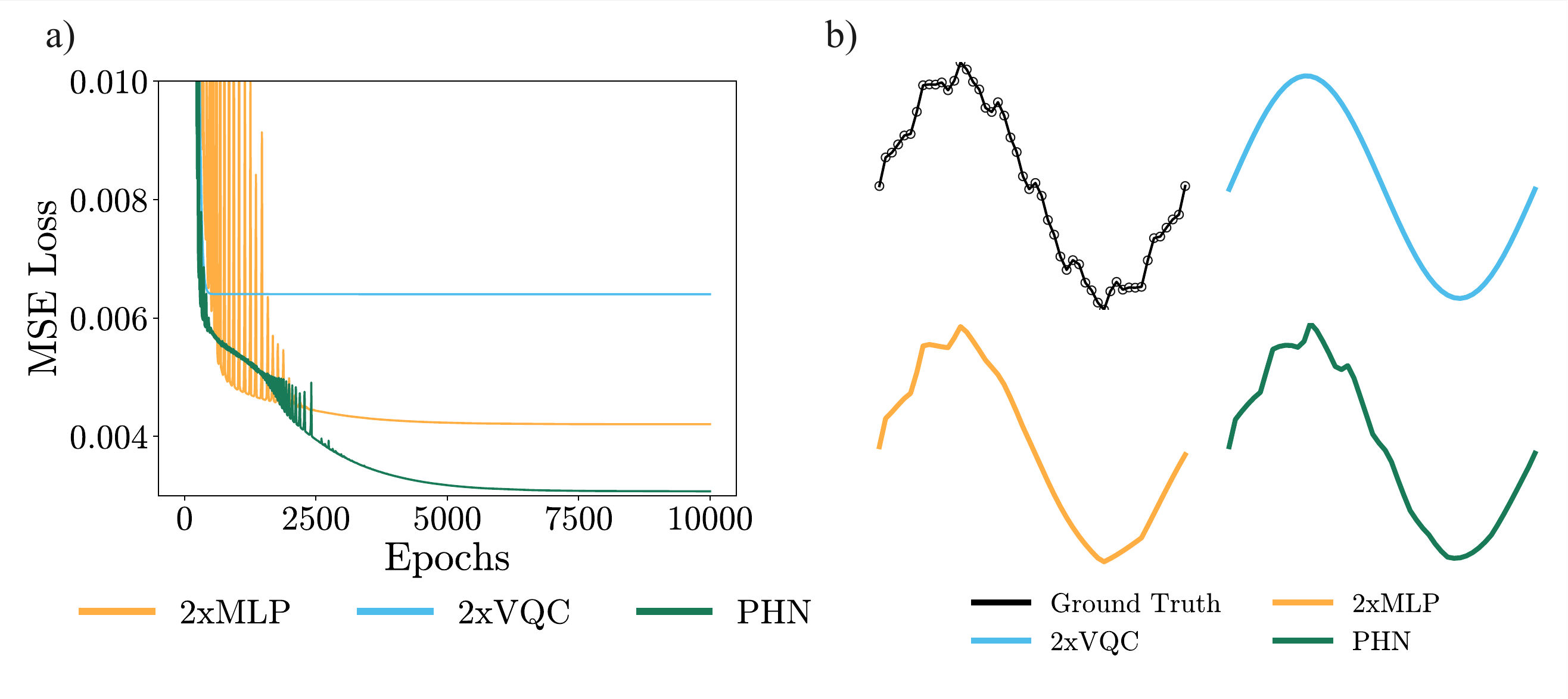}
    \caption{The results of the doubled networks when compared with the PHN. The PHN still outperforms the other architectures.}
    \label{fig:doubled}
\end{figure*}

This work compares the PHN with its individual constituents, the MLP and the VQC.  An important further experiment could parameterise the individual components to make a fairer comparison with the PHN. An intuitive way to create this is to stack two of each model, i.e. two VQCs in parallel or two MLPs in parallel, the doubled models.  Fig.~\ref{fig:doubled} shows the comparison of this setup on the one-dimensional dataset, where the PHN is compared with an MLP with doubled neurons and the VQC is stacked with another VQC of the same size.  Still, we observe that the PHN outperforms both models.

\section{Datasets with a large high-frequency term}\label{sec:high-freq}

\begin{figure*}[h]
    \centering
    \includegraphics[width=\textwidth]{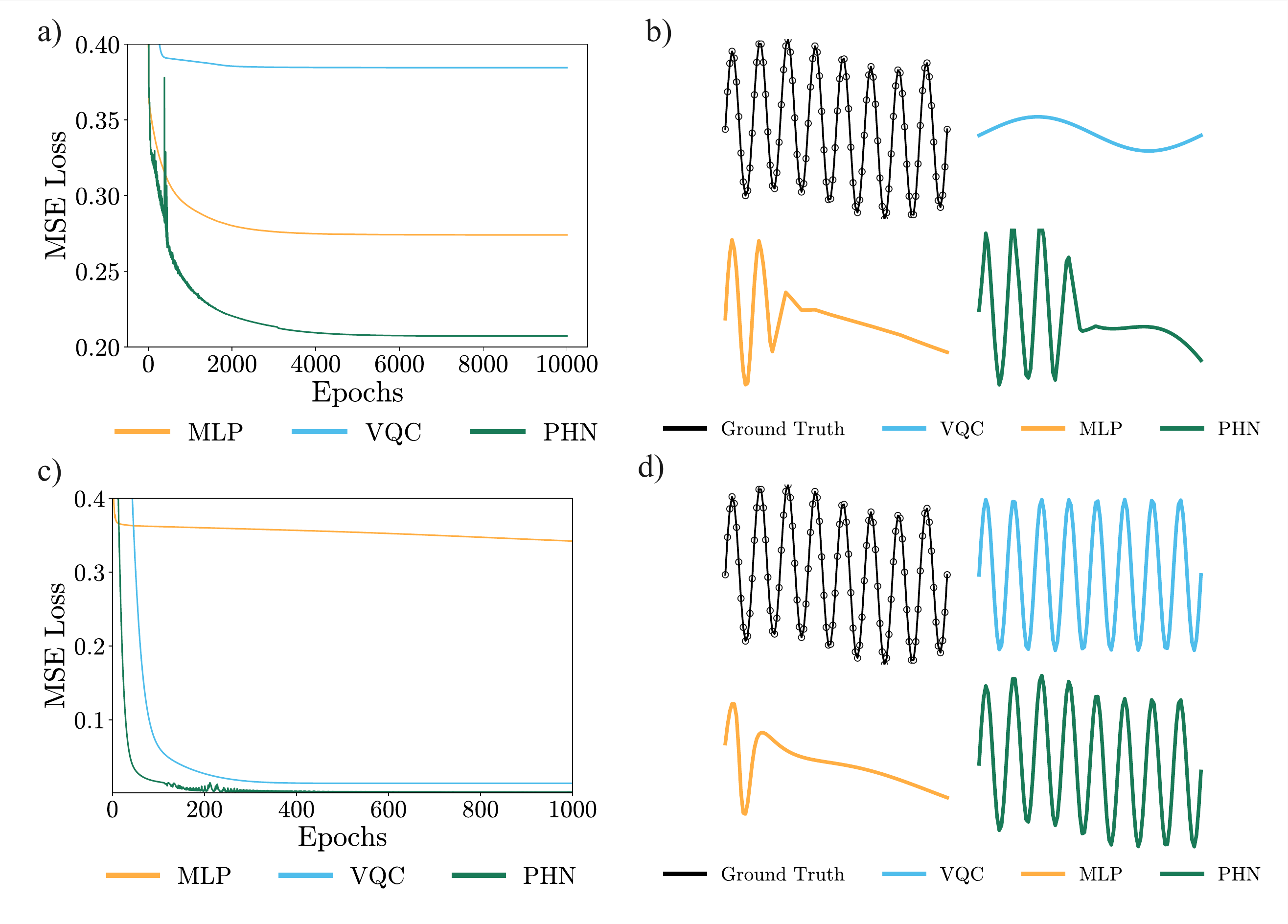}
    \caption{Illustration of the Parallel Hybrid Network's performance on a function with a dominant high-frequency component. (a) Represents the loss function of the naive case where PHN, MLP, and VQC were directly applied to the dataset, displaying varying degrees of error magnitude for each method. (b) Depicts the function fit for the naive case, demonstrating the efficacy of each approach in accurately fitting the function's waveform. (c) Shows the loss function for the scaled case where feature scaling was implemented before applying the PHN, MLP, and VQC, again highlighting the varying degrees of error magnitude. (d) Illustrates the function fits for the scaled case, evidencing a marked improvement in fit accuracy for the PHN and VQC post-scaling. This figure underscores the potential advantage of employing a PHN for such cases, outperforming other networks considered in this study.}
    \label{fig:large_high_frequency}
\end{figure*}

To extend the evaluation of the PHN's performance, a set of experiments was conducted on a function characterized by a dominant high-frequency component. The function employed for this analysis is given by $f(x)= 0.2 \sin(x) + \sin(8x)$. The problem was tackled through two approaches: 1) direct application of the PHN on the dataset, henceforth referred to as the naïve case, and 2) implementation of feature scaling prior to the application of the PHN. Fig.~\ref{fig:large_high_frequency} shows the losses and the fits of each of these approaches.

In the naïve case, it was observed that the VQC accurately identified the $\sin(x)$ term, although its expressivity was confined to a single Fourier term of frequency 1. The MLP accurately fitted the initial peaks but exhibited a downward trend beyond those points. Interestingly, the PHN provided the best fit, likely leveraging the classical component to conform to the high-frequency waves and the VQC to adhere to the underlying lower-frequency term. This behaviour exemplifies the promise held by the PHN as a suitable starting point for data analysis in scenarios with limited a-priori knowledge about the data.

For the case involving feature scaling, where the inputs were scaled by a factor of 8, a substantial improvement was recorded in the predictions of the VQC and the PHN, while the performance of the MLP was slightly degraded. This demonstrates that such preprocessing methods may not always confer advantages across all network types. The VQC, with its newfound capability to identify the $\sin(8x)$ term, highlights the potency of feature scaling in unlocking the full potential of VQCs. Meanwhile, the PHN also found a superior fit for the lower frequency term, outstripping the performances of both the MLP and the VQC.

These results accentuate the proficiency of the PHN in handling functions with dominant high-frequency components, particularly when coupled with feature scaling. Furthermore, they underscore the potential value of adopting a PHN for such cases, with its performance outpacing other networks considered in this study.

\section{Generalisation}\label{sec:generalisation}

A further investigation was carried out to test the generalisation ability of the PHN. The one-dimensional dataset was bifurcated, training the models only on the positive half and testing their ability to generalise to the untrained, negative half.

\begin{figure*}[h]
    \centering
    \includegraphics[width=\textwidth]{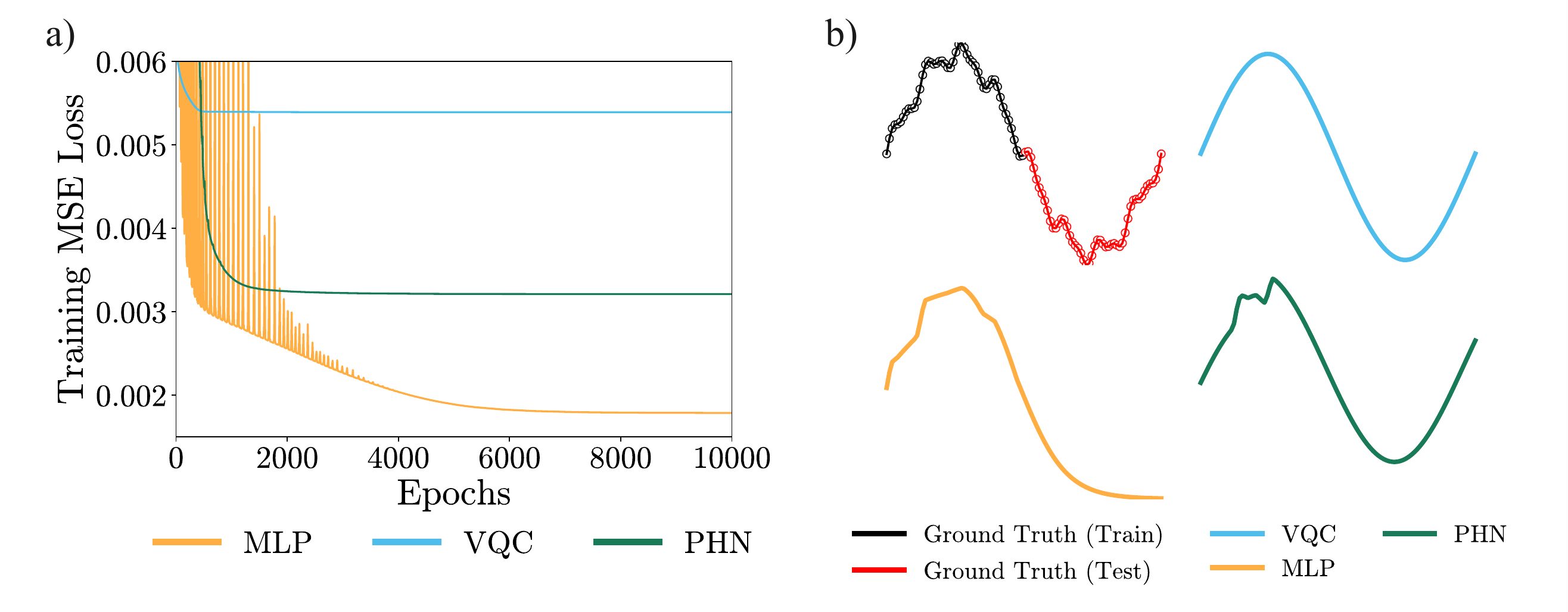}
    \caption{Comparative evaluation of extrapolation capabilities of the VQC, the PHN, and the MLP. The models were trained on the positive half of a one-dimensional dataset and then tested on the untrained negative half - see (a). The solid lines represent the learned functions of the respective models, with the MLP showing a significant deviation from the true function (dotted line) in the untrained region, indicating its limited extrapolation capability. In contrast, the VQC and the PHN demonstrate superior extrapolation capabilities, fitting the function well even in the untrained region due to their inherent ability to handle periodic functions. }
    \label{fig:generalisation}
\end{figure*}

The MLP successfully fitted the training data yet failed to predict behaviour external to this range, as visible in Fig.~\ref{fig:generalisation}. In the context of periodic functions, a default generalisation capability is observed for the Variational Quantum Classifier (VQC) due to the periodic nature of its special unitary members employed as linear mappings.

The PHN likewise inherits this extrapolation ability from the VQC, although no observed improvement in generalisation beyond the capability of the VQC is evident in this case.

The final test losses for the three models are as follows: for the VQC, the loss is measured at 0.008; for the PHN, the loss is recorded as 0.094; and for the MLP, the loss is computed as 0.133.

These findings demonstrate the superior extrapolation capability of the VQC and the PHN, particularly in the context of periodic functions, while highlighting the limitations of the MLP in this regard. Furthermore, it is suggested that although the PHN inherits the extrapolation capabilities of the VQC, it does not offer any additional benefit in terms of extrapolation in this particular case.

\end{document}